\documentclass[preprint,1p]{elsarticle}
\usepackage{amssymb}
\usepackage{lineno}
\usepackage{hyperref}
\usepackage[percent]{overpic}
\usepackage{enumitem}

\newcommand{\nq[1]}{\cdot 10^{#1} \; n_{eq}/cm^2 }

\begin{document}

\begin{frontmatter}

\title{Tracking particles at fluences  5-10$\cdot \mathbf {10^{16} \; n_{eq}/cm^2}$}

\author[1]{N.~Cartiglia\corref{cor}}\ead{cartiglia@to.infn.it}
\author[2]{H-F W.~Sadrozinski}
\author[2]{A.~Seiden}

\address[1]{INFN, Torino, Italy}
\address[2]{SCIPP, University of California Santa Cruz, CA, USA}

\cortext[cor]{Corresponding author}

\begin{abstract}
This paper presents the possibility of using very thin Low Gain Avalanche Diodes (LGAD) ($25 - 50\mu$m thick) as tracking detector at future hadron colliders, where particle fluence will be above $1\nq[16]$. In the present design, silicon sensors at the High-Luminosity LHC will be 100- 200 $\mu$m thick, generating, before irradiation, signals of 1-2 fC. This contribution  shows how very thin LGAD can provide signals of the same magnitude via the interplay of gain in the gain layer and gain in the bulk up to fluences above  $1\nq[16]$: up to fluences of 0.1-0.3$\nq[16]$, thin LGADs maintain a gain of $\sim$ 5-10 while at higher fluences the increased bias voltage will trigger the onset of multiplication in the bulk, providing the same gain as previously obtained in the gain layer. Key to this idea is the possibility of a reliable, high-density LGAD design able to hold large bias voltages ($\sim$ 500V).
\end{abstract}

\end{frontmatter}

\section{ Properties of Silicon sensors exposed to fluences above $1\cdot 10^{15} \; n_{eq}/cm^2 $ }
In the past several years, a lot of effort has been devoted to the study of the properties of Silicon sensors irradiated with fluences up to $1-2\nq[15]$. 
The extrapolation  of these results to fluences above $1\nq[15]$  depicted a very difficult situation: very high leakage currents, strong decline of charge collection efficiency, and a steep increase of the bulk doping. However, during the extensive experimental campaigns aimed at the development of the Silicon trackers to be operated at HL-LHC, it was found that  a simple linear extrapolation does not predict  accurately the situation above $1-5\nq[15]$: charge collection does not decrease due to trapping linearly with fluence and it remains still fairly high (above 50\%), detectors can hold very high biases (almost 1000V) allowing for charge multiplication to compensate charge trapping, the bulk doping does not increase linearly with fluence, and the leakage current increase is reduced. Overall, several preliminary measurements are suggesting that damage in Silicon sensors  does not increase linearly for fluences above $\sim 5\nq[15]$, however, sensors that are thicker than 50-100 microns still suffer from high leakage current, distortion of the electric field, charge trapping, and the impossibility of reaching full depletion. A comprehensive review on the effect of radiation damage  in Silicon can be found in ~\cite{HSTD11GK}.

\begin{figure}[htb]
\begin{center}
\includegraphics[width=.8\textwidth]{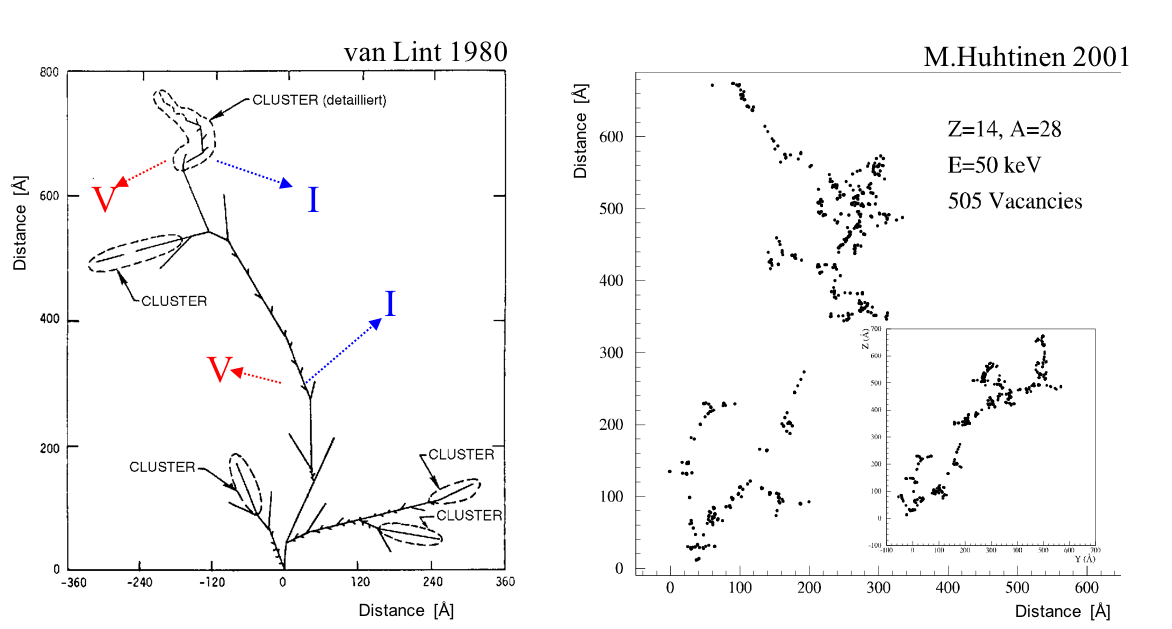}
\caption{Spatial distribution of the damaged produced by a 1 MeV neutron in Silicon. }
\label{fig:damage}
\end{center}
\end{figure}

From a phenomenological point of view,  the non linearity of the damage with fluence is expected: at high enough fluences, impinging particles will start hitting areas of Silicon that have already been hit previously and the resulting damage will happen on already damaged Silicon.  The geometrical distribution of the damage created by a particle in Silicon is fairly complex: Figure~\ref{fig:damage} ~\cite{HUHTINEN2002194, 1980STIA} shows two examples of the effects of a 1 MeV neutron. An interacting neutron creates several clusters of damaged Silicon,  with interstitial (I) or vacancy states (V); each cluster extends several tens of Angstroms, much more than the Silicon lattice constant  (5.4 \AA).

\begin{figure}[htb]
\begin{center}
\includegraphics[width=1.\textwidth]{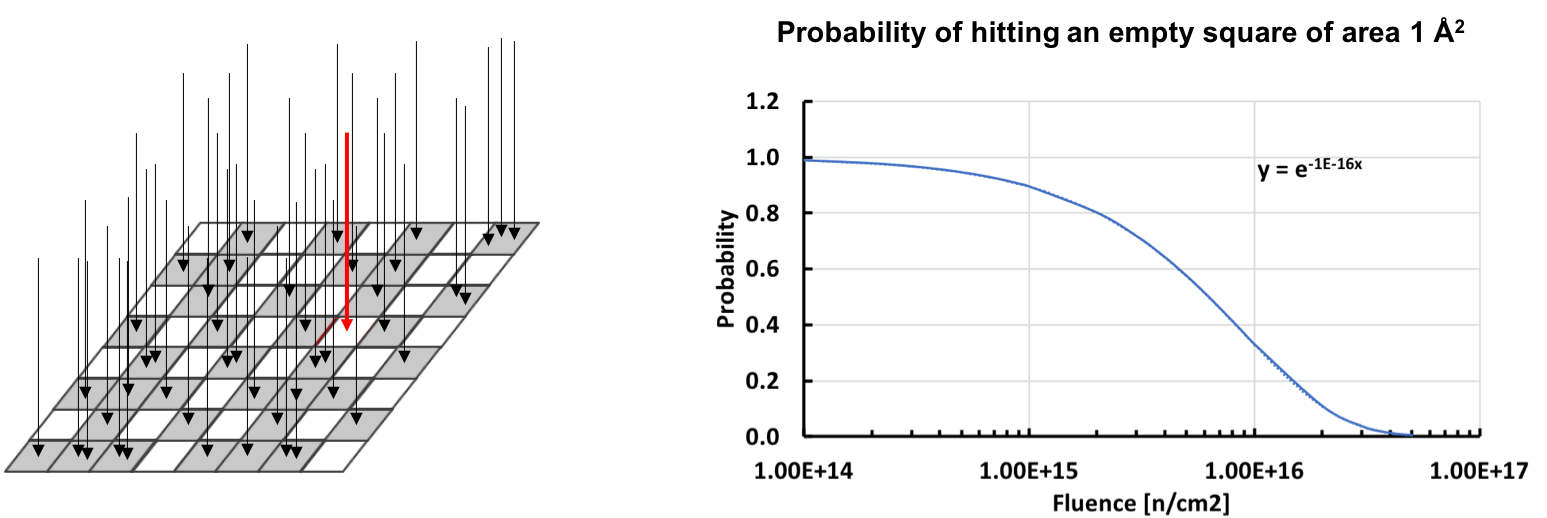}
\caption{Probability as a function of particle fluence of hitting a square of 1 \AA$^2$ not previously hit by any other particle. As the probability decreases, saturation effects in Silicon damage are more likely.}
\label{fig:cluster}
\end{center}
\end{figure}

The complete calculation of the  probability of  clusters overlap as a function of fluence is beyond the scope of this contribution, however, a simpler 2-dimensional approach can be used to gain insights into this problem.  The probability for an impinging particle on one cm$^2$ of Silicon to hit a location not already hit by  any of the preceding particles  can be calculated in the following manner,  Figure~\ref{fig:cluster} left side:
\begin{itemize}
\item A  particle hit on the  Silicon surface is identified by a square , $a_o$, of a given area, for example  1 \AA$^2$.   
\item For a particle, the probability of imping on a specific square is $ p_{hit} = \frac{10^{-16} cm^2}{cm^2} = 10^{-16}$.
\item Consequently, the probability of not-impinging on a specific square is $p_{miss} = 1 - p_{hit} = (1.-10^{-16})$
\item The probability for a particle to hit a square $a_o$ missed by all n previous particles is $p^n_{miss} = (p_{miss})^n = (1.-10^{-16})^n$.
\end{itemize}
The resulting trend is shown in Figure~\ref{fig:cluster} right side: for  $a_o$ =  1 \AA$^2$, after a fluence of $1\nq[16]$ the probability of hitting an empty square is reduced to 30\%, indicating that saturation effects are likely.  As shown in the plot, the probability of hitting an empty square as a function of fluence follows an exponential trend  with parameter $a_o$ since the events follow a Poisson distribution.

According to this simple model, the transition from linear to saturation happens in one decade of fluence. Measurements have shown that up to fluences $\sim 1-3 \nq[15]$ the damage is linear therefore the exploration of the properties of irradiated Silicon in the decade $ 3-30\nq[15]$ is of major importance to shed light on this topic.

\section{Thin Silicon sensors}

Even though saturation effects might lead to  better than foreseen properties of Silicon sensors at fluences above $ 5\nq[15]$, it is clear that the decrease of the carriers lifetime, the increase of leakage current and that of the bulk doping will severely impact operation. One way to minimize these negative effects is the use of thin p-bulk (25 - 50 $\mu$m) Silicon sensors: 
\begin{itemize}
\item High leakage current is responsible for noise increase and the distortion of the electric field due to the trapping of the charge carriers, forming the so-called  "double junction".  Thin sensors minimize this problem: as shown in ~\cite{HSTD11GK}, the first 50 $\mu$m of Silicon bulk near the n-p junction maintains  a linear electric field up to $ 1\nq[16]$,
\item The increase of bulk doping with fluence raises steeply the depletion voltage in 200-300 $\mu$m thick Silicon sensors: after $ 1\nq[15]$ the depletion voltage in a 300 (200) $\mu$m thick Silicon sensor is V $\sim 1400$V (620V).  Conversely, thin sensors can be depleted even after very high fluences: Figure ~\ref{fig:VFD} shows the depletion voltage after  a fluence of $ 1\nq[17]$ as a function of sensor thickness assuming a standard $ g_{eff}  = 0.02$ or a saturated $ g_{eff}  = 0.01$ value of the acceptor creation coefficient (for additional comments on the meaning of $g_{eff} $ see eq.\ref{eq:ac}  and  the discussion in chapter 5 of \cite{Balbuena:1291631}).  
\begin{figure}[htb]
\begin{center}
\includegraphics[width=0.7\textwidth]{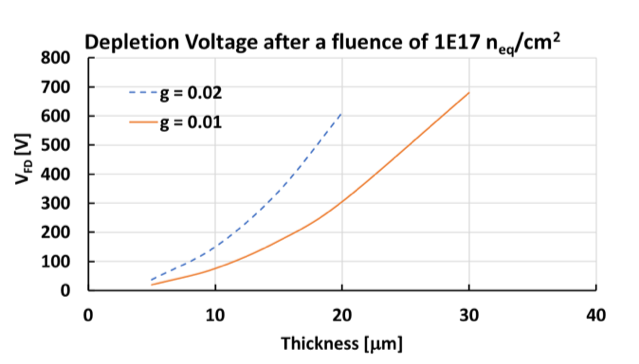}
\caption{Depletion voltage after  a fluence of $ 1\nq[17]$ as a function of sensor thickness assuming a standard $ g = 0.02$ or a saturated $ g = 0.01$ value of the acceptor creation coefficient.}
\label{fig:VFD}
\end{center}
\end{figure}
It is very important for reliable operation that the voltage of full depletion can be reached since it assure always a constant active volume and a field high enough to provide good carriers velocity everywhere in the volume
\item In thin sensors, even  if the carriers lifetimes becomes very short, charge collection efficiency remains fairly high: assuming an electron (hole) lifetime of 0.2 ns (0.15 ns) as predicted to be after a fluence of  $\sim 1\nq[16]$, charge collection efficiency in a  50 $\mu$m  thick sensor is still almost 80\%.
\end{itemize}

\begin{figure}[htb]
\begin{center}
\includegraphics[width=0.7\textwidth]{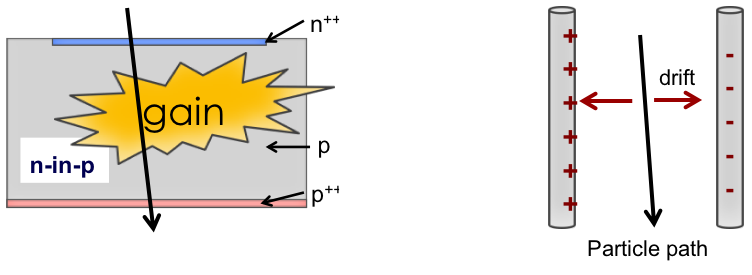}
\caption{Thin sensors with internal gain and 3D sensors manage to break the proportionality between drift path and signal amplitude.}
\label{fig:ig}
\end{center}
\end{figure}

Regardless of the advantages listed above, thin sensors (20 - 30 $\mu$m) are not suitable for operation since the signal is too small: the  most probable value of charge released by  a  minimum ionizing particle is of the order of ~ 0.3 fC while the newest ASICs developed by the RD53 collaboration\footnote{http://rd53.web.cern.ch/rd53/}, see for example ~\cite{TREDI2019_LD}, require at least 1 fC of charge to detect a  hit. For this reason, thin sensors can only be employed if they have an internal mechanism of charge multiplication.  As shown in Figure~\ref{fig:ig}, 3D sensors manage to break the proportionality between drift path and signal amplitude  by drifting the charge carriers perpendicularly to the direction of charge deposition: internal gain in thin sensors manages to  achieve the same results by multiplying the charge carriers. 

\section{Electric field configurations able to achieve internal gain}
\label{sec:ef}
\begin{figure}[htb]
\begin{center}
\includegraphics[width=1\textwidth]{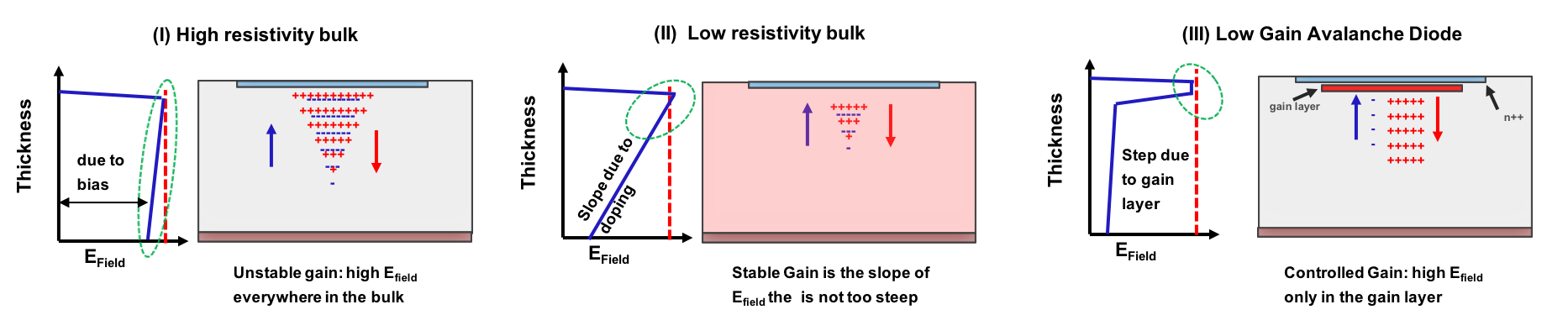}
\caption{The picture shows the electric fields of 3 different sensors: high resistivity, low resistivity and Low gain avalanche diode. Multiplication happens when the field is near the critical value (depicted as a dashed line). The sensors are n-in-p. The tones of red indicate the level of acceptor doping in both bulk and gain layer.}
\label{fig:cm}
\end{center}
\end{figure}

The internal gain in Silicon detectors, often called impact ionization,  happens when the electric field reaches values around 250-300 kV/cm~\cite{ROPP}. The gain as a function of position $x$ is calculated as the path integral of the charges drifting towards the electrodes: 

\begin{equation}
\label{eq:g}
N(x) = \int N(o) e^{\alpha(E) x} dx, 
\end{equation}

where $\alpha(E)^{-1}$ is the mean path needed by a charge carrier to acquire enough velocity to create an additional e/h pair. This distance decreases exponentially with the electric field $E$ ~\cite{ROPP}. In a simplified picture, charge multiplication happens if the electrons drift length  is longer than $\alpha(E)^{-1}$: lower values of $E$ require very long drift distances while high values of $E$ field (250-300kV/cm) require $\sim 100$ nm. As shown in  Figure~\ref{fig:cm},   there are 3 methods to implement a high enough field to induce charge multiplication:

\begin{itemize}
\item (I) If the bulk  has high resistivity, the field is mostly due to the applied bias. In this circumstance, the electric field is almost flat so it is either always or never near the critical field (shown in the picture as a dashed line). Sensors in this configuration tend to be very unstable, as a small variation of the bias value or of the temperature  can stop or ignite a breakdown. 
\item (II) The electric field in sensors with a low resistivity bulk has a slope proportional to the bulk doping: multiplication happens  near the junction since it is where the field is higher.  The problem with this configuration is that the electric field increases constantly approaching the junction, therefore it can easily generates an avalanche at its highest point. For this reason, stable gain is achieved only  if the slope of the field is not too steep.
\item (III) In the Low Gain Avalanche Diode~\cite{LGAD1} design (LGAD), the gain layer doping  produces a sudden increase of the $E$ field  at a  distance of 1-2 microns from the junction. The total field,  due to the sum of the gain layer and the bias components,  is near the critical field only for a short distance, it is flat,  and it is controlled by the bias value. The combination of these three factors allows having a controlled gain.  This configuration offers the most control over the gain as the field is flat, and near the critical field only for a short length. 
\end{itemize} 

The above examples identify the main conditions for stable gain: (i) the field should be near the critical value only for a short distance (less than a few microns), (ii) the field near  the critical value should be as flat as possible, and (iii) the field value should be controlled by bias and not by bulk doping.  For these reasons, only the LGAD design can provide the internal gain necessary for the stable operation of thin sensors. 

\section{Irradiation effects}
\label{sec:ie}
Irradiation modifies the doping profile of the bulk and of the gain layer  by progressively  deactivating the  initial acceptors (in the bulk and in the  gain layer) and by  creating acceptor-like defects due to deep traps. These effects are described by  equation (\ref{eq:ac})~\cite{ROPP, Fer}

\begin{equation}
\label{eq:ac}
N_A(\phi) = g_{eff} \phi + N_A(0) e^{-c\phi}, 
\end{equation}

where $g_{eff}$ = 0.02  [cm$^{-1}$],  $\phi$ the irradiation fluence   [ cm$^{-2}$],  $(N_{A}(0))$ $(N_{A}(\phi))$ the accceptor density [cm$^-3$] before irradiation (after fluence $\phi$).  the initial (after a fluence $\phi$) acceptor density [cm$^{-3}$], and $c$ [cm$^2$] is a constant  that depends on the initial acceptor concentration and on the type of irradiation. 
The first term of equation (\ref{eq:ac}) accounts for acceptor creation by deep traps  while the second term for the initial acceptor removal mechanism.  As a result of increasing fluence, the gain layer becomes less and less doped while the bulk doping increases: this evolution decreases the gain in the gain layer and, at high enough fluence and bias voltage, might generate gain in the bulk. 

\begin{figure}[htb]
\begin{center}
\includegraphics[width=1\textwidth]{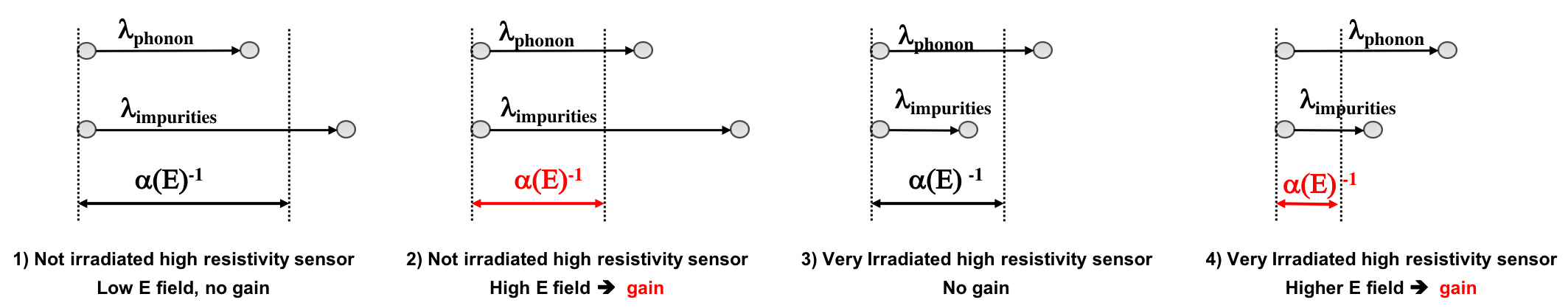}
\caption{Interplay of the scattering length $\lambda$ and the mean path needed for impact ionization $\alpha(E)^{-1}$.}
\label{fig:la}
\end{center}
\end{figure}

\vspace{0.3cm}
One key question to be addressed in future R\&D is the dependence of the gain mechanism on irradiation. 
In un-irradiated high-resistivity sensors, the charge carriers mean free path $\lambda$ between successive scatterings is determined by the interactions of the carriers with the lattice phonons  while the scattering on impurities  is a sub-leading effect. Under this condition, at high enough $E$ field  $\alpha(E)^{-1}$ becomes shorter than  $\lambda$ and the process of impact ionization can take place,  see the first two left sketches of Figure~\ref{fig:la}. Irradiation increases the number of scattering centers and at high enough fluence the distance $\lambda$ will be controlled by the scattering on defects: when this will happen the gain will be quenched, and it would only be recovered  by increasing the $E$ field as shown in the  two right sketches of Figure~\ref{fig:la}.  Current studies have proven that up to fluences $~3\nq[15]$ impact ionization is not quenched, indicating that the scattering on phonons is still the limiting factor.  

\section{Design of the sensors able to work at fluences $5-10\cdot 10^{16} \; n_{eq}/cm^2 $ }

\begin{figure}[htb]
\begin{center}
\includegraphics[width=1\textwidth]{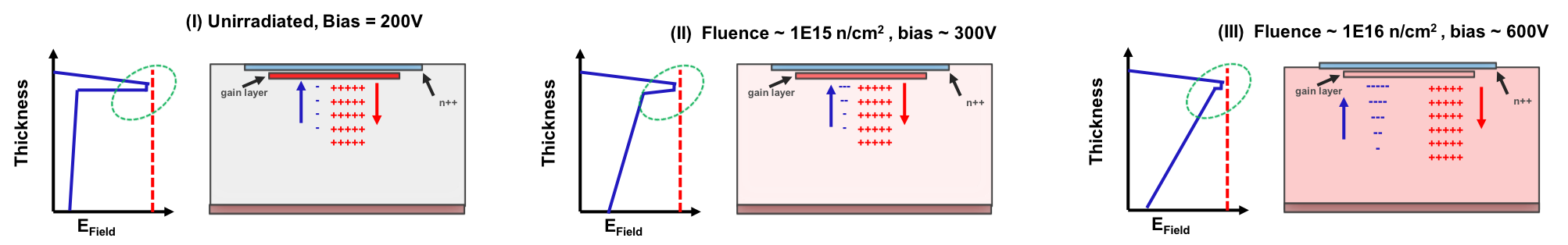}
\caption{The picture shows the electric fields of a thin LGAD when (i) new , (ii) after a fluence of  $1\nq[15]$ and (III) after $1\nq[16]$. When new, the field is mostly due to the gain layer, while with irradiation the bulk becomes more doped and the gain layer disappears. As a consequence of this change, the multiplications moves from the gain layer to the bulk.  The sensors are n-in-p, the tones of red indicated the level of acceptor doping both in the bulk and the gain layer.}
\label{fig:icm}
\end{center}
\end{figure}

The combination of the results obtained in Section~\ref{sec:ef} with the irradiation effects explained in  Section~\ref{sec:ie} suggest that a thin LGAD device will be able to deliver enough charge up to fluences of $5-10\nq[16]$: at low fluence the gain is due to the  presence of a gain layer while at higher fluence the gain is  due to multiplication in the bulk, see Figure~\ref{fig:icm}. Multiplication in the bulk is driven by the bulk doping and by the bias voltage, and it will be present up to fluences  where the mean free path $\lambda$ will be too short to allow $\alpha(E)^{-1} < \lambda$.  Since up to fluences of $\sim 3\nq[15]$ impact ionization is not quenched,  that the damage might not increase linearly,  with fluence, and that the $E$ field in thin sensors can be very high, the possibilities of having gain at fluences in the interval $5-10\nq[16]$ looks possible. 

It is important that the sensors can be  over-depleted even at the highest fluence so that the field remains as flat as possible when it approaches the critical value and does not lead to uncontrollable breakdown.  According to Figure~\ref{fig:VFD} a thickness of about 20-25 microns would be ideal as it can still be depleted even after a fluence of $1\nq[17]$ and the initial charge is large enough that a moderate value of gain, gain = 5 - 10, will be sufficient to guarantee the delivery of at least 1 fC of charge at every fluence.

\begin{figure}[htb]
\begin{center}
\includegraphics[width=1.\textwidth]{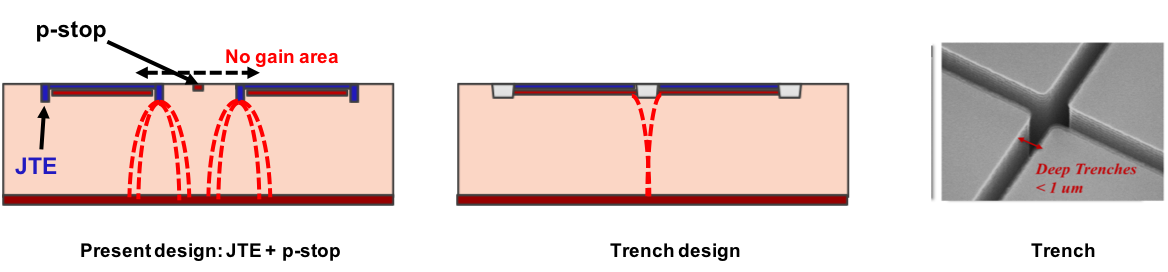}
\caption{Left side: In the current LGAD design, the gain is terminated using junction termination extensions (JTE) and the n-doped pixels are isolated from each other using p-stops. This design creates a dead area between pads. In order to eliminate the dead area it has been proposed to use trenches  to terminate the gain and isolate the pads. }
\label{fig:tr}
\end{center}
\end{figure}

An important aspect  in the design of Silicon sensors  is that the fill factor, defined as the active area divided by the sensor area, should be as close to 100\% as possible. In the LGAD traditional design,  Figure~\ref{fig:tr} left part, the gain is terminated using junction termination extensions (JTE) and the n-doped pixels are isolated from each other using p-stops: these structures use at least 30$\mu$m, strongly decreasing the fill factor. To circumvent this problem, it has been proposed~\cite{GP-RD50-H} to use shallow trenches  to terminate the gain and isolate the pads,   Figure~\ref{fig:tr} right part. If successful, this design change will be able to dramatically improve the fill factor opening up the possibility of designing very small LGAD (50x50 $\mu m^2$) pixels.

\section{Conclusion}
\begin{figure}[htb]
\begin{center}
\includegraphics[width=0.7\textwidth]{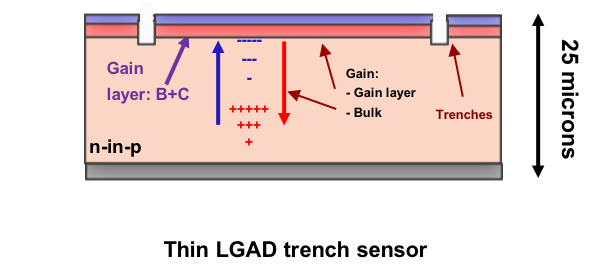}
\caption{Sketch of the proposed sensor for particle tracking at very high fluences.}
\label{fig:fi}
\end{center}
\end{figure}

This contribution proposes the use of thin LGAD  as tracking sensors for very high fluences. Thin devices minimize irradiation effects, greatly reducing the effect of charge trapping, high leakage current, and increasing bulk doping.  Thin LGADs boost the initially small signal using the internal gain: the interplay of charge multiplication in the gain layer and in the bulk will assure the deliver of more than 1fC of charge even after fluences of $5-10\nq[16]$. An innovative design  for the interpad area based on shallow trenches will allow the production of  LGAD with 100\% fill factor. A summary sketch of the proposed design is shown in Figure~\ref{fig:fi}.

\section{Acknowledgments}
Part of this work has been financed by the European Union's Horizon 2020 Research and Innovation funding program, under Grant Agreement no. 654168 (AIDA-2020) and Grant Agreement no. 669529 (ERC UFSD669529), and by the Italian Ministero degli Affari Esteri and INFN Gruppo V. The work was supported by the United States Department of Energy, grant  DE-FG02-04ER41286.

\bibliography{trk17_HEP}

\end{document}